\documentclass[twocolumn,aps,pre,amsmath]{revtex4-1}
\usepackage[final]{graphics}

\DeclareMathOperator{\Tr}{Tr}

\begin{document}

\title{Statistical mechanics of homogeneous partly pinned fluid
  systems}

\author{Vincent Krakoviack}

\affiliation{Laboratoire de Chimie, {\'E}cole Normale Sup{\'e}rieure
  de Lyon, 46 All{\'e}e d'Italie, 69364 Lyon Cedex 07, France}

\date{\today}

\begin{abstract}
  The homogeneous partly pinned fluid systems are simple models of a
  fluid confined in a disordered porous matrix obtained by arresting
  randomly chosen particles in a one-component bulk fluid or one of
  the two components of a binary mixture. In this paper, their
  configurational properties are investigated. It is shown that a
  peculiar complementarity exists between the mobile and immobile
  phases, which originates from the fact that the solid is prepared in
  presence of and in equilibrium with the adsorbed fluid.  Simple
  identities follow, which connect different types of configurational
  averages, either relative to the fluid-matrix system or to the bulk
  fluid from which it is prepared.  Crucial simplifications result for
  the computation of important structural quantities, both in computer
  simulations and in theoretical approaches. Finally, possible
  applications of the model in the field of dynamics in confinement or
  in strongly asymmetric mixtures are suggested.
\end{abstract}

\maketitle

\section{Introduction}

The physics of fluids under nanoscale confinement is a topic of great
interest, from a fundamental, applied and interdisciplinary point of
view at once (see
\cite{GelGubRadSli99RPP,AlcMcK05JPCM,AlbCoaDosDudGubRadSli06JPCM} for
reviews). This is however a rather difficult one for
theorists. Indeed, since many porous solids are structurally
disordered, one is usually faced with a complex interplay of finite
size effects, adsorption phenomena, topology, and randomness, which is
not easily captured with analytically or numerically tractable models.

In the past few years, a widely used class of models for theoretical
and computational studies of fluids adsorbed in disordered porous
media has been one in which the fluid molecules evolve in a
statistically homogeneous random array of particles frozen in a
disordered configuration sampled from a prescribed probability
distribution. Recent works include investigations of the
non-equilibrium phenomena in the adsorption/desorption processes
\cite{KieRosTarVio01PCCP,KieMonRosSarTar01PRL,KieMonRosTar02JPCM,%
  RosKieTar03EL}, of the critical behavior at fluid-fluid transitions
\cite{VinBinLow06PRL,VinBinLow08JPCM,PelVinCacLow08JPCM,%
  DeSPel08PRL,Vin09SM}, and of the glassy dynamics in confinement
\cite{GalPelRov02EL,GalPelRov03aPRE,GalPelRov03bPRE,%
  AttGalRov05JCP,GalAttRov09PRE,Kim03EL,KimMiySai09EL,%
  Kra05PRL,Kra05JPCM,Kra07PRE,Kra09PRE,KurCosKah09PRL}.

The first elaborate theoretical treatment of this type of systems has
been derived by Madden and Glandt \cite{MadGla88JSP} for the model of
the so-called ``quenched-annealed'' (QA) binary mixture. In this
model, the probability distribution of the porous matrix is chosen as
the equilibrium distribution of some simple fluid system, so that the
solid samples can be thought of as the results of instantaneous
thermal quenches of this original equilibrium system, hence the
denomination ``quenched'' for the matrix component. Then the fluid
(the ``annealed'' component) equilibrates in the random potential
energy landscape created by the frozen matrix particles.  Thanks to
the property of statistical homogeneity of the solid, Madden and
Glandt have demonstrated that the QA mixture can be studied with great
ease via simple and direct extensions of standard liquid-state
theoretical methods. In their pioneering and subsequent works
\cite{MadGla88JSP,FanGlaMad90JCP,Mad92JCP,ForGla94JCP}, they used
diagrammatic techniques to investigate the distribution functions and
thermodynamic properties of the system. Later, Stell and coworkers
rederived, extended, and, in some cases, corrected these results with
the help of the replica trick
\cite{GivSte92JCP,LomGivSteWeiLev93PRE,GivSte94PA,RosTarSte94JCP}.
Interesting developments in the framework of the density functional
theory have been reported as well
\cite{Sch02PRE,Sch05JPCM,Sch09PRE,LafCue06PRE}.

Following this breakthrough, other prescriptions have been put forward
for the preparation of the disordered porous matrix, resulting in a
variety of models. In a depleted system
\cite{TasTalVioTar97PRE,Tas97JCP}, the matrix is obtained by quenching
configurations of a one-component fluid and by removing at random a
fraction of its particles. Strategies for a correlated depletion step
have recently been suggested \cite{ZhaDonLiu07JCP,ZhaDonLiu07JML}. In
a templated system \cite{Tas99PRE,ZhaTas00JCP,ZhaTas00MP,%
  ZhaCheTas01PRE}, the porous medium is produced by freezing
configurations of a binary mixture and by removing one of its
components, which acts as a template for the remaining one. In both
models, the probability distribution of the solid can be explicitly
and easily related to the equilibrium distribution of an underlying
simple fluid system, as in the QA mixture. This needs not be the
case. Indeed, for many aspects of the formalism, it has been shown
that it is enough to know the structure of the solid, typically at the
pair level, instead of its full statistics
\cite{Mad92JCP,Mad95JCPcom}. This has allowed theoretical
investigations of fluids confined in realistic porous media generated
through out-of-equilibrium processes, such as the diffusion-limited
cluster-cluster aggregation (DLCA) process \cite{KraKieRosTar01JCP}.

In all the above examples, the preparation of the matrix and thus its
statistical properties are independent of the fluid that will be
imbibed in it. This looks like the most reasonable way to proceed,
since this actually reproduces the usual experimental
situation. However, a rather popular model in computer simulation
studies of confined fluids does not display this feature
\cite{VirAraMed95PRL,VirMedAra95PRE,ChaJuaMed08PRE,Kim03EL,%
  KimMiySai09EL,ChaJagYet04PRE,MitErrTru06PRE,FenMryPryFol09PRE}. This
model is the randomly pinned (RP) system, which is considered in the
present paper with one of its natural generalizations. Operationally,
it can be described as follows. While exploring the configuration
space of an equilibrium bulk fluid system, snapshots are selected at
random times, in which a fraction of the particles is chosen randomly
and pinned down, thus generating samples of rigid disordered
matrices. The remaining unpinned particles then become the confined
fluid phase which evolves under the influence of the same interactions
as in the original bulk system. Note that a very similar setup can be
realized experimentally in two dimensions, by squeezing a binary
colloidal mixture between two glass plates
\cite{CruSauAra98PRL,CruAra99PRE}.

It will be shown in the following that the RP model is actually a
special case of a templated system. As such, its core physics does not
display any fundamental difference compared to any other type of
fluid-matrix model. But, in this specific setup, it is also clear that
the fluid and matrix properties are unusually and irremediably
entangled. In fact, since the porous matrix is prepared in the
presence of the fluid, there is a very peculiar complementarity
between the mobile and frozen phases, from which we will show that
nontrivial configurational properties emerge which make the RP model
worth special attention. Furthermore, among the different
particle-based models of disordered porous media, the RP system is
unique in allowing of an interpretation as a limiting case of an
equilibrated binary mixture, whereby the mass of the pinned particles
is sent to infinity (for Newtonian dynamics) or their free-diffusion
coefficient to zero (for Brownian dynamics)
\cite{VirAraMed95PRL,VirMedAra95PRE,ChaJuaMed08PRE,%
  FenMryPryFol09PRE}.  Its study is thus relevant for the
understanding of the dynamics of mixtures with a strong dynamical
asymmetry, of which it represents an asymptote.

One of the properties that will be established in the present paper
has already been observed in the above-cited computer simulation
studies, namely that the pair distribution functions of the bulk
system on which the RP model is based are preserved after the pinning
process
\cite{VirAraMed95PRL,VirMedAra95PRE,Kim03EL,ChaJagYet04PRE}. Up to now
and depending on the authors, however, it was never clear whether this
finding was reflecting an exact property of the system or was just a
good approximation. The issue is settled here in favor of the first
interpretation. In addition, it will be shown how other distribution
functions, which are characteristic of systems with quenched disorder
and usually quite difficult to compute efficiently, should be easily
accessible in the very special case of the RP model.

The paper is organized as follows. In Sec.~II, the RP model and a
useful generalization are described with their defining probability
distributions. In Sec.~III, configurational identities are derived,
which relate different types of averages. They are the main results of
this work, whose possible applications in computer simulation studies
are discussed. Their consequences for the structure at the pair level
and their implications for integral equation theories are considered
in Sec.~IV, while a dynamical point of view is developed in
Sec.~V. Section~VI is devoted to concluding remarks.

\section{Models and probability distributions}

In this section, the RP model is defined and its essential statistical
properties are derived. To achieve this goal, we proceed in two
steps. First, a formal link is established between the RP and
templated systems, which leads to a natural generalization of the
problem. Second, the relevant probability distributions are provided
in this extended framework.

\begin{figure*}
  \includegraphics*{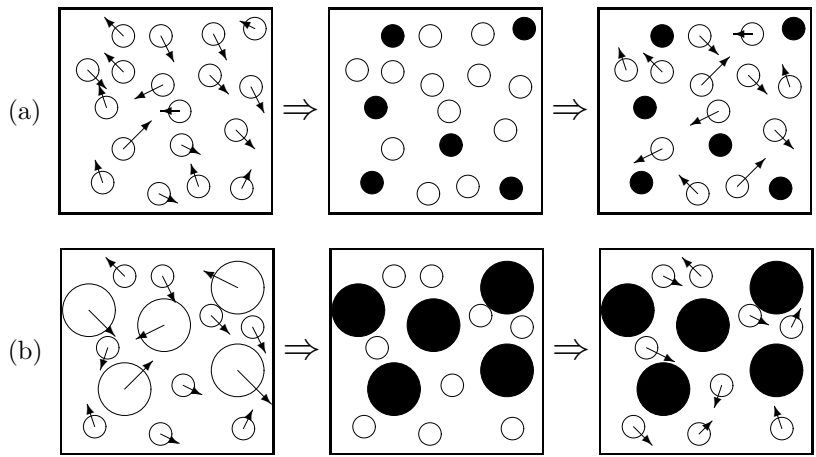}
  \caption{\label{figsketch} Simple schematics of the fluid-matrix
    models studied in this work. In all panels, the immobile
    particles are represented in black, and the mobile ones in white,
    with arrows symbolizing their movement. (a) Randomly pinned
    system. In a precursor one-component fluid (left panel), a
    predefined fraction of the particles is chosen at random in an
    instantaneous configuration and pinned down to form the matrix
    (middle panel, the pinning fraction is $x=1/3$), in which the
    remaining particles continue their motion (right panel). (b)
    Partly pinned system. Starting with an equilibrated binary mixture
    (left panel), one component is pinned down in an instantaneous
    configuration to form the matrix (middle panel), in which the
    other component continues its movement (right panel).}
\end{figure*}

The simplest version of the RP model, which is sufficient to
illustrate the main properties of this type of systems, derives from a
one-component bulk fluid through a random pinning process, as sketched
in Fig.~\ref{figsketch} \footnote{Multicomponent versions are possible
  for which every species has its own pinning fraction. The
  generalization of the results of the present work to this case is
  straightforward.}. More specifically, for each configuration of the
fluid which occurs with a probability distribution corresponding to a
prescribed statistical ensemble, a fraction $x$ of its particles is
randomly chosen and pinned down to form a disordered porous matrix,
while the complementary $1-x$ fraction remains mobile and provides the
confined fluid. Different ensembles can be considered, which are all
equivalent in the thermodynamic limit \cite{RosTarSte94JCP}. In
existing computer simulation studies
\cite{VirAraMed95PRL,VirMedAra95PRE,%
  ChaJuaMed08PRE,Kim03EL,KimMiySai09EL,ChaJagYet04PRE,MitErrTru06PRE,%
  FenMryPryFol09PRE}, both the fluid and the matrix have always been
treated in the canonical ensemble, i.e., with no fluctuations in their
particle numbers and thus in the pinning fraction. Here, we find more
convenient to resort to a grand-canonical description in which, in
particular, the pinning fraction is allowed to fluctuate around its
mean value.

Accordingly, we start with a one-component bulk fluid in a volume $V$
at temperature $T$ (as usual, we define $\beta = 1/k_\text{B} T$) and
activity $z$. From well-known statistical mechanics
\cite{macdohansen3ed}, the configurational probability density of
finding this system with $N$ particles located at $(\mathbf{r}_1,
\mathbf{r}_2, \ldots, \mathbf{r}_N) \equiv \mathbf{r}^{N}$ is
\begin{equation}\label{pbulk}
  \mathcal{P}_\text{bulk}(N,\mathbf{r}^{N}) = \frac{z^{N}  e^{ -\beta
      V_\text{bulk}(N,\mathbf{r}^{N})}}{\Xi_\text{bulk} N!},
\end{equation}
where $V_\text{bulk}(N,\mathbf{r}^{N})$ denotes the potential energy
of the fluid. The partition function $\Xi_\text{bulk}$ is simply
\begin{equation}
  \Xi_\text{bulk} = \Tr \frac{z^{N} e^{-\beta V_\text{bulk}(N,
      \mathbf{r}^{N})}}{N!},
\end{equation}
where the shorthand notation for the trace operator $\Tr \cdots \equiv
\sum_{N=0}^{+\infty} \int d\mathbf{r}^N \cdots$ is used.

During the pinning process, for every configuration of the fluid, each
particle in the system might be pinned down with probability $x$ or
left mobile with probability $1-x$.  It results that the joint
probability density of generating a matrix with $N_\text{m}$ immobile
particles located at $(\mathbf{q}_1, \mathbf{q}_2, \ldots,
\mathbf{q}_{N_\text{m}}) \equiv \mathbf{q}^{N_\text{m}}$ while
$N_\text{t}$ unpinned particles are located at $(\mathbf{s}_1,
\mathbf{s}_2, \ldots, \mathbf{s}_{N_\text{t}}) \equiv
\mathbf{s}^{N_\text{t}} $ at the time of the pinning process, with
$N_\text{m}+N_\text{t} = N$, is
\begin{multline}\label{pinjoint}
  \mathcal{P}_\text{mt}(N_\text{m},\mathbf{q}^{N_\text{m}},
  N_\text{t},\mathbf{s}^{N_\text{t}}) = \\
  \frac{z^{N_\text{m}+N_\text{t}} x^{N_\text{m}} (1-x)^{N_\text{t}}
    e^{ -\beta V_\text{bulk}(N_\text{m}+N_\text{t},
      \mathbf{q}^{N_\text{m}},
      \mathbf{s}^{N_\text{t}})}}{\Xi_\text{bulk} N_\text{m}!
    N_\text{t}!},
\end{multline}
which is basically Eq.~\eqref{pbulk} modified by a combinatorial
factor due to the random pinning process. For clarity, different
notations are used for the locations of the pinned and unpinned
particles, and, in anticipation of the coming interpretation, the
subscripts $\text{m}$ and $\text{t}$ for ``matrix'' and ``template'',
respectively, have been introduced.  The probability density of the
matrix configurations immediately follows by tracing out the unpinned
particles,
\begin{equation} \label{pinmatrix}
  \mathcal{P}_\text{m}(N_\text{m},\mathbf{q}^{N_\text{m}}) =
  \Tr_\text{t} \mathcal{P}_\text{mt}(N_\text{m},\mathbf{q}^{N_\text{m}},
  N_\text{t},\mathbf{s}^{N_\text{t}}). 
\end{equation}
This is exactly what would be obtained for a depleted system
\cite{TasTalVioTar97PRE,Tas97JCP}. This is not surprising, since it is
rather intuitive that a RP system is equivalent to a depleted system
in which the depleted particles are reinjected as the confined fluid.

A similar equivalence can be established with a templated system
\cite{Tas99PRE,ZhaTas00JCP,ZhaTas00MP,ZhaCheTas01PRE}. Indeed, thanks
to the binomial identity, $\Xi_\text{bulk}$ can be rewritten as
\begin{multline}\label{pinpartition}
  \Xi_\text{bulk} = \Tr_\text{m} \Tr_\text{t} \\
  \frac{z^{N_\text{m}+N_\text{t}} x^{N_\text{m}} (1-x)^{N_\text{t}}
    e^{ -\beta V_\text{bulk}(N_\text{m}+N_\text{t},
      \mathbf{q}^{N_\text{m}}, \mathbf{s}^{N_\text{t}})}} {N_\text{m}!
    N_\text{t}!}.
\end{multline}
$\mathcal{P}_\text{mt} (N_\text{m}, \mathbf{q}^{N_\text{m}},
N_\text{t}, \mathbf{s}^{N_\text{t}})$ is then easily recognized as the
grand-canonical configurational probability density of an ideal binary
mixture, with activities $z_\text{m} = zx$ and $z_\text{t} = z(1-x)$
for the components m and t, respectively, and $\mathcal{P}_\text{m}
(N_\text{m}, \mathbf{q}^{N_\text{m}})$ as the probability distribution
of the templated matrices that can be generated from this ideal
mixture, components m and t being the matrix and template components,
respectively. Note that $V_\text{bulk}$ appears unchanged, because the
potential energy of an ideal binary mixture is, by definition,
independent of its composition and thus equal to that of any one of
its components alone. Therefore, a RP system with a pinning fraction
$x$ is equivalent to a templated system based on an ideal binary
mixture with a matrix number fraction $x$, in which the template
particles are reinjected as the mobile phase.

An immediate generalization of the problem follows from this
equivalence, by relaxing the restriction to ideal mixtures (see
Fig.~\ref{figsketch}). We call the corresponding extended class of
models the partly pinned (PP) systems, which might be studied
generically and only specialized afterwards to deal with the RP
model. Thus, instead of Eqs.~\eqref{pinjoint} and
\eqref{pinpartition}, we shall consider the generic expressions for
templated systems, i.e.,
\begin{multline}\label{joint}
  \mathcal{P}_\text{mt}(N_\text{m},\mathbf{q}^{N_\text{m}},
  N_\text{t},\mathbf{s}^{N_\text{t}}) = \\
  \frac{z_\text{m}^{N_\text{m}} z_\text{t}^{N_\text{t}} e^{ -\beta
      \left[ V_\text{mm}(N_\text{m}, \mathbf{q}^{N_\text{m}}) +
        V_\text{mt}(N_\text{m}, \mathbf{q}^{N_\text{m}}, N_\text{t},
        \mathbf{s}^{N_\text{t}}) + V_\text{tt}(N_\text{t},
        \mathbf{s}^{N_\text{t}}) \right] }} {\Xi_\text{bulk} N_\text{m}!
    N_\text{t}!}
\end{multline}
for the grand-canonical configurational probability distribution of
the bulk matrix-template system and for the joint probability density
of generating a matrix with $N_\text{m}$ particles located at
$\mathbf{q}^{N_\text{m}}$ while $N_\text{t}$ template particles are
located at $\mathbf{s}^{N_\text{t}}$, with
\begin{multline}\label{partition}
  \Xi_\text{bulk} = \Tr_\text{m} \Tr_\text{t} \\
  \frac{z_\text{m}^{N_\text{m}} z_\text{t}^{N_\text{t}} e^{ -\beta
      \left[ V_\text{mm}(N_\text{m}, \mathbf{q}^{N_\text{m}}) +
        V_\text{mt}(N_\text{m}, \mathbf{q}^{N_\text{m}}, N_\text{t},
        \mathbf{s}^{N_\text{t}}) + V_\text{tt}(N_\text{t},
        \mathbf{s}^{N_\text{t}}) \right] }}{N_\text{m}!  N_\text{t}!}
\end{multline}
for the normalizing partition sum.  Equation~\eqref{pinmatrix} still
generates the matrix probability distribution.  In the above
equations, $z_\text{m}$ and $z_\text{t}$ are the activities of the
matrix and template components in the bulk binary mixture, and
$V_\text{mm}(N_\text{m}, \mathbf{q}^{N_\text{m}})$,
$V_\text{mt}(N_\text{m}, \mathbf{q}^{N_\text{m}}, N_\text{t},
\mathbf{s}^{N_\text{t}})$ and $V_\text{tt}(N_\text{t},
\mathbf{s}^{N_\text{t}})$ are the matrix-matrix, matrix-template and
template-template contributions to the potential energy of the
mixture, respectively.

So far, only probabilities relative to the matrix and its preparation
process have been considered. We now turn to the confined fluid
statistics. As mentioned above, the confined fluid is assumed to be in
grand-canonical equilibrium in the presence of the porous matrix and,
by construction of the model, to inherit its properties from the
template. Thus, its activity $z_\text{f}$ (the subscript $\text{f}$ is
used for properties of the confined fluid) is taken equal to the
template activity $z_\text{t}$ and, for a particular realization of
the matrix with $N_\text{m}$ particles located at
$\mathbf{q}^{N_\text{m}}$, its potential energy is chosen as
$V_\text{mt}(N_\text{m}, \mathbf{q}^{N_\text{m}}, N_\text{f},
\mathbf{r}^{N_\text{f}}) + V_\text{tt}(N_\text{f},
\mathbf{r}^{N_\text{f}})$, when $N_\text{f}$ fluid particles located
at $(\mathbf{r}_1, \mathbf{r}_2, \ldots, \mathbf{r}_{N_\text{f}})
\equiv \mathbf{r}^{N_\text{f}}$ are present in the system. It follows
that the probability density of such a configuration, which depends
parametrically on the matrix configuration
$(N_\text{m},\mathbf{q}^{N_\text{m}})$, is
\begin{multline} \label{fluidprob}
  \mathcal{P}_\text{f}(N_\text{f},\mathbf{r}^{N_\text{f}} |
  N_\text{m},\mathbf{q}^{N_\text{m}}) = \\
  \frac{z_\text{t}^{N_\text{f}} e^{ -\beta \left[
        V_\text{mt}(N_\text{m}, \mathbf{q}^{N_\text{m}}, N_\text{f},
        \mathbf{r}^{N_\text{f}}) + V_\text{tt}(N_\text{f},
        \mathbf{r}^{N_\text{f}}) \right]
    }}{\Xi_\text{f}(N_\text{m},\mathbf{q}^{N_\text{m}}) N_\text{f}!}
\end{multline}
with the confined fluid partition function
\begin{multline}
  \Xi_\text{f}(N_\text{m},\mathbf{q}^{N_\text{m}}) = \\ \Tr_\text{f}
  \frac{z_\text{t}^{N_\text{f}} e^{ -\beta \left[
        V_\text{mt}(N_\text{m}, \mathbf{q}^{N_\text{m}}, N_\text{f},
        \mathbf{r}^{N_\text{f}}) + V_\text{tt}(N_\text{f},
        \mathbf{r}^{N_\text{f}}) \right]}}{N_\text{f}!}.
\end{multline}

Combining Eqs.~\eqref{joint} and \eqref{fluidprob}, a very simple
equality can be derived,
\begin{multline} \label{identity} \mathcal{P}_\text{mt}(N_\text{m},
  \mathbf{q}^{N_\text{m}}, N_\text{t}, \mathbf{s}^{N_\text{t}})
  \mathcal{P}_\text{f}(N_\text{f}, \mathbf{r}^{N_\text{f}} |
  N_\text{m}, \mathbf{q}^{N_\text{m}}) = \\
  \mathcal{P}_\text{mt}(N_\text{m}, \mathbf{q}^{N_\text{m}},
  N_\text{f},\mathbf{r}^{N_\text{f}} )
  \mathcal{P}_\text{f}(N_\text{t},\mathbf{s}^{N_\text{t}} |
  N_\text{m},\mathbf{q}^{N_\text{m}}),
\end{multline}
which reflects the very peculiar symmetries of the system under study
and will play a crucial role in the following. 

Armed with these results, we might now investigate the configurational
properties of the RP and PP systems.

\section{Configurational identities} \label{identities}

Dealing with quenched-disordered fluid-matrix systems, two types of
configurational averages have to be considered when computing their
properties \cite{MadGla88JSP,Mad92JCP}. The first one is the ordinary
thermal average denoted by $\langle \cdots \rangle$, taken for a given
realization $(N_\text{m}, \mathbf{q}^{N_\text{m}})$ of the matrix with
the probability density $\mathcal{P}_\text{f}(N_\text{f},
\mathbf{r}^{N_\text{f}} | N_\text{m},\mathbf{q}^{N_\text{m}})$. The
second one is the disorder average over the matrix realizations,
denoted by $\overline{\cdots}$, to be taken with the probability
density $\mathcal{P}_\text{m}(N_\text{m},\mathbf{q}^{N_\text{m}})$
after the thermal average. We shall not dwell on subtleties of the
physics of quenched-disordered systems, but it might be worth
recalling that this average over disorder is equivalent to an average
over macroscopic subparts of a macroscopic system for additive
quantities, and to a volume average over a macroscopic sample for
locally defined quantities \cite{LifGrePasbook}.

These two types of averages might be combined in many different
ways. For instance, computations of free energy differences typically
involve expressions of the form $\overline{\ln \langle A
  \rangle}$. Here, we shall concentrate on two specific examples for
which simple results can be derived. They correspond to double
averages $\overline{\langle A \rangle}$ and to products of the form
$\overline{\langle A \rangle \langle B \rangle}$. Such quantities are
often combined to generate correlation functions characterizing the
two physically distinct types of fluctuations present in disordered
systems. Thus, the typical thermal fluctuations are quantified by
so-called connected averages of the form $\overline{\langle A B
  \rangle} - \overline{\langle A \rangle \langle B \rangle}$, while
the disorder-induced fluctuations of thermal quantities are measured
by so-called disconnected averages defined as $\overline{\langle A
  \rangle \langle B \rangle} - \overline{\langle A \rangle}\
\overline{\langle B \rangle}$. The distinction between these two types
of correlation functions is an essential feature of the physics of
disordered systems, whose significance is, for instance, stressed by
the fact that the thermodynamic susceptibilities, such as the
isothermal compressibility \cite{ForGla94JCP,RosTarSte94JCP}, are
always expressed as connected averages.

In order to shorten the equations in this rather formal section, a
condensed vector notation will be used. Thus, we define $\mathbf{m}
\equiv (N_\text{m}, \mathbf{q}^{N_\text{m}})$, $\mathbf{t} \equiv
(N_\text{t},\mathbf{s}^{N_\text{t}})$, and $\mathbf{f} \equiv
(N_\text{f},\mathbf{r}^{N_\text{f}})$. With these definitions,
Eq.~\eqref{identity} now reads
\begin{equation}\label{identitybis} 
  \mathcal{P}_\text{mt}(\mathbf{m},\mathbf{t})
  \mathcal{P}_\text{f}(\mathbf{f}|\mathbf{m}) = 
  \mathcal{P}_\text{mt}(\mathbf{m},\mathbf{f})
  \mathcal{P}_\text{f}(\mathbf{t}|\mathbf{m}).
\end{equation}

Generically, a configurational variable for a quenched-disordered
fluid-matrix system is a function
$A(\mathbf{x};\mathbf{m},\mathbf{f})$ of the matrix and fluid particle
numbers and coordinates, possibly with other variables (such as space
variables when dealing with $n$-particle densities) collectively
denoted by $\mathbf{x}$. Its thermal average for a given matrix
realization is defined as
\begin{equation}
  \langle A(\mathbf{x};\mathbf{m},\mathbf{f}) \rangle = 
  \Tr_\text{f} \mathcal{P}_\text{f}(\mathbf{f}|\mathbf{m})
  A(\mathbf{x};\mathbf{m},\mathbf{f}),
\end{equation}
so that its double average after tracing out the matrix variables
reads
\begin{equation}
  \overline{\langle A(\mathbf{x};\mathbf{m},\mathbf{f}) \rangle} =
  \Tr_\text{m} \mathcal{P}_\text{m}(\mathbf{m}) \Tr_\text{f}
  \mathcal{P}_\text{f}(\mathbf{f}|\mathbf{m}) 
  A(\mathbf{x};\mathbf{m},\mathbf{f}).
\end{equation}
Specializing to the PP systems and introducing an explicit reference
to the original matrix-template mixture, this can be rewritten as
\begin{equation}
  \overline{\langle A(\mathbf{x};\mathbf{m},\mathbf{f}) \rangle} =
  \Tr_\text{m} \Tr_\text{t}
  \Tr_\text{f} 
  \mathcal{P}_\text{mt}(\mathbf{m},\mathbf{t})
  \mathcal{P}_\text{f}(\mathbf{f}|\mathbf{m}) 
  A(\mathbf{x};\mathbf{m},\mathbf{f}),
\end{equation}
which, thanks to Eq.~\eqref{identitybis}, is transformed into
\begin{equation}\label{factorized} 
  \overline{\langle A(\mathbf{x};\mathbf{m},\mathbf{f}) \rangle} =
  \Tr_\text{m} \Tr_\text{t} \Tr_\text{f}
  \mathcal{P}_\text{mt}(\mathbf{m},\mathbf{f}) 
  \mathcal{P}_\text{f}(\mathbf{t}|\mathbf{m})
  A(\mathbf{x};\mathbf{m},\mathbf{f}), 
\end{equation}
eventually leading to
\begin{equation}\label{mixtureaverage} 
  \overline{\langle A(\mathbf{x};\mathbf{m},\mathbf{f}) \rangle} =
  \Tr_\text{m} \Tr_\text{f}
  \mathcal{P}_\text{mt}(\mathbf{m},\mathbf{f})
  A(\mathbf{x};\mathbf{m},\mathbf{f}), 
\end{equation}
since $\Tr_\text{t} \mathcal{P}_\text{f}(\mathbf{t}|\mathbf{m}) = 1$
appears factorized in Eq.~\eqref{factorized}.

The right-hand side of Eq.~\eqref{mixtureaverage} is immediately
recognized as a thermal average for the bulk matrix-template mixture,
which will be denoted by $\langle \cdots \rangle_\text{bulk}$. So, the
first sought-for configurational identity reads
\begin{equation}\label{firstidentity}
  \overline{\langle A(\mathbf{x};\mathbf{m},\mathbf{f}) \rangle} =
  \langle A(\mathbf{x};\mathbf{m},\mathbf{f}) \rangle_\text{bulk},
\end{equation}
i.e., the double-averaged configurational properties of a PP system
coincide with the corresponding quantities in the bulk fluid on which
it is based \footnote{A similar result for functions of the fluid
  variables only has been reported for heterogeneous partly pinned
  systems in Ref.~\cite{SchKobBin04JPCB}}. This in particular applies
to the $n$-particle densities and distribution functions and explains
past observations in computer simulation studies
\cite{VirAraMed95PRL,VirMedAra95PRE,Kim03EL,ChaJagYet04PRE}.

There are different possible ways of taking advantage of this result
in computer simulation studies. An immediate idea is that one can
completely avoid the computation of double averages for the PP
fluid-matrix systems. Instead, the corresponding calculations can be
performed on the bulk fluid, which in any case has to be simulated in
order to generate the porous samples and for which the problem is
conceptually simpler (only one type of average is required and it is
an ordinary thermal average) and the sampling of the configuration
space is often more efficient, thanks to faster dynamics
\cite{VirAraMed95PRL,VirMedAra95PRE,Kim03EL,ChaJagYet04PRE,%
  MitErrTru06PRE,FenMryPryFol09PRE}. Another option is to use the
computation of double averages as a means to calibrate the parameters
of the simulation study. Indeed, the above identity is not expected to
hold on a sample-by-sample basis. It becomes valid only after the
disorder average is performed. Therefore, one can use the comparison
between double averages and high quality data for the bulk as a guide
to estimate the minimal number of matrix realizations that is required
in order to achieve a satisfactory convergence of the disorder
averaging procedure.

In the case of two configurational variables
$A(\mathbf{x};\mathbf{m},\mathbf{f})$ and
$B(\mathbf{y};\mathbf{m},\mathbf{f})$, the typical value of the
product of their thermal averages is defined as
\begin{multline}
  \overline{\langle A(\mathbf{x};\mathbf{m},\mathbf{f}) \rangle
    \langle B(\mathbf{y};\mathbf{m},\mathbf{f'}) \rangle} =
  \Tr_\text{m} \mathcal{P}_\text{m}(\mathbf{m}) \\
  \Tr_\text{f} \mathcal{P}_\text{f}(\mathbf{f}|\mathbf{m})
  A(\mathbf{x};\mathbf{m},\mathbf{f}) \Tr'_\text{f}
  \mathcal{P}_\text{f}(\mathbf{f'}|\mathbf{m})
  B(\mathbf{y};\mathbf{m},\mathbf{f'}),
\end{multline}
where $\Tr'_\text{f}$ simply represents the trace over primed fluid
variables.  Through Eq.~\eqref{pinmatrix}, this might be rewritten for
a PP system as
\begin{multline}
  \overline{\langle A(\mathbf{x};\mathbf{m},\mathbf{f}) \rangle
    \langle B(\mathbf{y};\mathbf{m},\mathbf{f'}) \rangle} =
  \Tr_\text{m} \Tr_\text{t} \Tr_\text{f} \Tr'_\text{f}
  \mathcal{P}_\text{mt}(\mathbf{m},\mathbf{t}) \\
  \mathcal{P}_\text{f}(\mathbf{f}|\mathbf{m})
  A(\mathbf{x};\mathbf{m},\mathbf{f})
  \mathcal{P}_\text{f}(\mathbf{f'}|\mathbf{m})
  B(\mathbf{y};\mathbf{m},\mathbf{f'}).
\end{multline}
Thanks to Eq.~\eqref{identitybis}, one can exchange the variables
$\mathbf{t}$ and $\mathbf{f'}$ between
$\mathcal{P}_\text{mt}(\mathbf{m},\mathbf{t})$ and
$\mathcal{P}_\text{f}(\mathbf{f'}|\mathbf{m})$ (one could choose to
exchange $\mathbf{t}$ and $\mathbf{f}$ instead, but this would simply
amount to an exchange of $A$ and $B$ which play symmetric roles in the
original problem), then perform the trace $\Tr_\text{t}$ which reduces
to a normalization condition as above. It remains
\begin{multline}
  \overline{\langle A(\mathbf{x};\mathbf{m},\mathbf{f}) \rangle
    \langle B(\mathbf{y};\mathbf{m},\mathbf{f'}) \rangle} =
  \Tr_\text{m} \Tr_\text{f} \Tr'_\text{f} \\
  \mathcal{P}_\text{mt}(\mathbf{m},\mathbf{f'})
  \mathcal{P}_\text{f}(\mathbf{f}|\mathbf{m})
  A(\mathbf{x};\mathbf{m},\mathbf{f})
  B(\mathbf{y};\mathbf{m},\mathbf{f'}),
\end{multline}
whose meaning is illuminated by a simple change of dummy variables
leading to
\begin{multline}\label{templateaverage} 
  \overline{\langle A(\mathbf{x};\mathbf{m},\mathbf{f}) \rangle
    \langle B(\mathbf{y};\mathbf{m},\mathbf{f'}) \rangle} = 
  \Tr_\text{m} \Tr_\text{t} \Tr_\text{f} \\
  \mathcal{P}_\text{mt}(\mathbf{m},\mathbf{t})
  \mathcal{P}_\text{f}(\mathbf{f}|\mathbf{m})
  A(\mathbf{x};\mathbf{m},\mathbf{f})
  B(\mathbf{y};\mathbf{m},\mathbf{t}).
\end{multline}

The right-hand side of the latter equation takes the form of a double
average, but now with a modified disorder average involving the
matrix-template probability distribution
$\mathcal{P}_\text{mt}(\mathbf{m},\mathbf{t})$, which will be denoted
by $\overline{\cdots}'$ in the following. Such a modification is very
natural if one includes the template variables $(N_\text{t},
\mathbf{s}^{N_\text{t}})$ in the set of configurational parameters
describing the system and accordingly deals with configurational
variables of the form
$A(\mathbf{x};\mathbf{m},\mathbf{t},\mathbf{f})$. Strictly speaking,
this extension is not required by the physics of the system, which can
be discussed uniquely in terms of fluid and matrix parameters
\cite{Mad92JCP}, in which case, as it should be, the modified disorder
averaging procedure does not change anything, as attested by
identities such as
\begin{equation}\label{primednonprimed} 
  \overline{\langle A(\mathbf{x};\mathbf{m},\mathbf{f}) \rangle}' =
  \overline{\langle A(\mathbf{x};\mathbf{m},\mathbf{f}) \rangle}.
\end{equation}
However, the consideration of template degrees of freedom offers a
very pictorial way to capture how the matrix-template and
template-template correlations imprinted in the porous solid during
its preparation are transferred to the fluid. It can also lead to
important technical simplifications in formal developments. One
example based on the interpretation of Eq.~\eqref{templateaverage}
will follow. Another one is provided by the integral equation theory
of the generic templated systems \cite{Tas99PRE,ZhaTas00JCP,%
  ZhaTas00MP,ZhaCheTas01PRE}, where the introduction of total and
direct correlation functions involving the template leads to equations
that preserve standard diagrammatic prescriptions (no nodal points in
direct correlation functions), at variance with Madden's more compact
formalism \cite{Mad92JCP}.

Coming back to Eq.~\eqref{templateaverage}, the second sought-for
configurational identity thus reads
\begin{equation}\label{secondidentity}
  \overline{\langle A(\mathbf{x};\mathbf{m},\mathbf{f}) \rangle
    \langle B(\mathbf{y};\mathbf{m},\mathbf{f'} \rangle} = 
  \overline{\langle A(\mathbf{x};\mathbf{m},\mathbf{f})
    B(\mathbf{y};\mathbf{m},\mathbf{t}) \rangle}',
\end{equation}
where, since the matrix and template variables are static quantities,
$B(\mathbf{y};\mathbf{m},\mathbf{t})$ can be placed inside or outside
the thermal average. So, in a PP system, the disorder-averaged product
of two thermal averages coincides with the modified double-averaged
product of a fluid-matrix configurational variable with a
template-matrix function.  Interestingly, this result can be
formulated in terms of non-equilibrium quantities as well. Indeed,
since the fluid and the template components are essentially identical,
the template coordinates at the time of preparation of the matrix can
be considered as initial conditions for an ulterior fluid dynamics
inside the matrix. So, the right-hand side of
Eq.~\eqref{secondidentity} can be interpreted as the correlation of a
thermal average with a function of these initial conditions.

The availability of Eq.~\eqref{secondidentity} represents a major
simplification for computational studies of PP systems compared to
other particle-based models of disordered porous media. Indeed, the
modified double average on its right-hand side should be no more
difficult to compute than the standard double average on fluid and
matrix configurations met in all this family of systems. The only
additional price to pay is to keep track of the positions occupied by
the template particles at the moment of the matrix production. Thus,
through Eq.~\eqref{secondidentity}, quantities of the form
$\overline{\langle A \rangle \langle B \rangle}$, which are required
to compute the physically important connected correlations, are made
numerically accessible via a simple and controlled procedure.

The relevance of this result is best illustrated through an example,
which is the computation of the two-point correlation function
\begin{equation}
  \psi(\mathbf{r},\mathbf{r}') = \overline{\langle
    \hat{\rho}^{(1)}_\text{f}(\mathbf{r};\mathbf{f}) \rangle \langle
    \hat{\rho}^{(1)}_\text{f}(\mathbf{r}';\mathbf{f'}) \rangle},
\end{equation}
where $\hat{\rho}^{(1)}_\text{f}(\mathbf{r};\mathbf{f})$ is the
microscopic fluid density operator
\begin{equation} \label{oneparticledensity}
  \hat{\rho}^{(1)}_\text{f}(\mathbf{r};\mathbf{f}) =
  \sum_{i=1}^{N_\text{f}} \delta(\mathbf{r} - \mathbf{r}_i).
\end{equation}
For a generic fluid-matrix system, one typically has to calculate
$\langle \hat{\rho}^{(1)}_\text{f}(\mathbf{r};\mathbf{f}) \rangle$ on
a grid for each matrix realization and to compute the autocorrelations
as sums over the grid points, before disorder averaging
\cite{MerLevWei96JCP}. Unfortunately, the outcome of this procedure
has been found to be marred by artifacts related to the finite size of
the grid cells. Analogous difficulties are present in reciprocal space
calculations of the structure factor associated with
$\psi(\mathbf{r},\mathbf{r}')$ \cite{SchCosKurKah09MP}. In the case of
a PP system, however, Eq.~\eqref{secondidentity} can be used to obtain
\begin{equation}\label{blockingdensity} 
  \psi(\mathbf{r},\mathbf{r}') = \overline{\langle
    \hat{\rho}^{(1)}_\text{f}(\mathbf{r};\mathbf{f}) 
    \hat{\rho}^{(1)}_\text{t}(\mathbf{r}';\mathbf{t}) \rangle}',
\end{equation}
where $\hat{\rho}^{(1)}_\text{t}(\mathbf{r};\mathbf{t})$ is the
template analogue of
$\hat{\rho}^{(1)}_\text{f}(\mathbf{r};\mathbf{f})$. Further
simplification occurs by noting that the product
$\hat{\rho}^{(1)}_\text{f}(\mathbf{r};\mathbf{f})
\hat{\rho}^{(1)}_\text{t}(\mathbf{r}';\mathbf{t})$ is simply the
two-particle fluid-template density operator
$\hat{\rho}^{(2)}_\text{ft}(\mathbf{r},\mathbf{r'};\mathbf{f},\mathbf{t})$,
so that eventually
\begin{equation} 
  \psi(\mathbf{r},\mathbf{r}') = \overline{\langle
    \hat{\rho}^{(2)}_\text{ft}(\mathbf{r},\mathbf{r'};\mathbf{f},\mathbf{t})
    \rangle}'.
\end{equation}
The computation of $\psi(\mathbf{r},\mathbf{r}')$ is thus turned into
the rather innocuous problem of computing the pair correlation
function between the fluid particles and the template sites, which
does not require the use of a grid and thus is obviously free of grid
artifacts. It is interesting to note that an analogous possibility to
replace the direct computation of a disconnected correlation function
by the computation of an equivalent pair correlation function also
exists for ideal gases in disordered environments, thanks to the fact
that the connected pair correlation functions identically vanish in
these systems. The case has been investigated on one specific example
in Ref.~\cite{MerLevWei96JCP}, where it is reported that the latter
strategy yields better results than the former. This clearly lends
support to the suggestion that the availability of
Eq.~\eqref{secondidentity} and the resulting simplifications for PP
systems should facilitate the accumulation of high quality computer
simulation data.

Finally, the above equations, formulated in the framework of the PP
systems, can be easily adapted to the case of the
one-component-fluid-based RP models. In particular, using
Eq.~\eqref{firstidentity}, the double-averaged quantities in the RP
system can be related to corresponding thermal averages in the
one-component bulk fluid. One only needs to revert the mapping to an
ideal template-matrix mixture described in the previous
section. Unfortunately, the simplicity of the above identities is lost
in the process, since the dependence of the resulting expressions on
the pinning fraction is found to change with the configurational
variable under consideration. For this reason, we do not report any
specific results for the RP model here. Some important ones will
appear in the next section.

\section{Pair correlations and integral equation theories} 

The description of the structure of fluids at the pair level is a
central issue of liquid state theory. In this section, we examine how
the above identities impact on this problem for the PP and RP
fluid-matrix models.

Since we are dealing with special cases of templated systems, the
natural framework for the present study is the theory developed by Van
Tassel and coworkers
\cite{Tas99PRE,ZhaTas00JCP,ZhaTas00MP,ZhaCheTas01PRE}.  For a generic
templated system with no special symmetries, they introduced eight
total correlation functions in order to fully describe the structure
at the pair level. Six of them are the standard functions reflecting
the two-body correlations between the different types of particles
involved in the model. They are the matrix-matrix, matrix-template,
template-template, matrix-fluid, template-fluid, and fluid-fluid total
correlation functions, denoted by $h_\text{mm}(r)$, $h_\text{mt}(r)$,
$h_\text{tt}(r)$, $h_\text{mf}(r)$, $h_\text{tf}(r)$, and
$h_\text{f\/f}(r)$, respectively. They can be expressed in terms of
double-averaged two-particle density operators $\hat{\rho}^{(2)}_{ij}
(\mathbf{r}, \mathbf{r'})$, $i,j=\text{m,t,f}$ (for notational
convenience, the dependence on the particle numbers and configurations
is omitted here and in the following), as
\begin{equation}
  h_{ij}(|\mathbf{r}-\mathbf{r'}|) = \frac{\overline{ \langle
      \hat{\rho}^{(2)}_{ij}(\mathbf{r}, \mathbf{r'}) \rangle}' -
    \rho_{i} \rho_{j}}{\rho_{i} \rho_{j}}, 
\end{equation}
where $\rho_{i} = \overline{\langle \hat{\rho}^{(1)}_{i} (\mathbf{r})
  \rangle}'$ is the number density of species $i$ [see
Eq.~\eqref{oneparticledensity} for the expression of
$\hat{\rho}^{(1)}_{i} (\mathbf{r})$]. Note that a modified (primed)
disorder average is used, because it is required for the proper
definition of quantities pertaining to template particles.  When only
matrix and/or fluid particles are involved, it can safely be replaced
by the usual matrix average on the basis of
Eq.~\eqref{primednonprimed}. Through these averages, the statistical
isotropy and homogeneity of the system are restored, hence the
dependence of all pair correlation functions on the scalar
$|\mathbf{r}-\mathbf{r'}|$ only.

The two remaining functions are specific to quenched-disordered
systems and give an account of the correlations between the one-body
fluid densities at two different points induced by the disorder. They
are the blocking or disconnected total correlation function
\begin{equation}
  h_\text{b}(|\mathbf{r}-\mathbf{r'}|) = \frac{\overline{ \langle
      \hat{\rho}^{(1)}_\text{f}(\mathbf{r}) \rangle \langle
      \hat{\rho}^{(1)}_\text{f}(\mathbf{r'}) \rangle} - \rho_\text{f}^2}
  {\rho_\text{f}^2},
\end{equation}
and the connected total correlation function
\begin{equation}
  h_\text{c}(|\mathbf{r}-\mathbf{r'}|) = \frac{\overline{ \langle
      \hat{\rho}^{(2)}_\text{f\/f}(\mathbf{r}, \mathbf{r'}) \rangle} -
    \overline{ \langle \hat{\rho}^{(1)}_\text{f}(\mathbf{r}) \rangle
      \langle \hat{\rho}^{(1)}_\text{f}(\mathbf{r'})
      \rangle}}{\rho_\text{f}^2}, 
\end{equation}
hence the discussion of the function $\psi(\mathbf{r},\mathbf{r}')$ in
the previous section. Note that $h_\text{f\/f}(r)$, $h_\text{b}(r)$ and
$h_\text{c}(r)$ are not independent, since they obey $h_\text{f\/f}(r) =
h_\text{b}(r) + h_\text{c}(r)$.

Using diagrammatic techniques or the replica trick, Ornstein-Zernike
(OZ) equations are easily derived which connect the above total
correlation functions to the corresponding set of direct correlation
functions, $c_{ij}(r)$, $c_\text{b}(r)$ and $c_\text{c}(r)$. They read
(for convenience, the $r$ dependence is omitted)
\begin{subequations}\begin{align}
  h_\text{mm} & = c_\text{mm} + \rho_\text{m} c_\text{mm} \otimes
  h_\text{mm} + \rho_\text{t} c_\text{mt} \otimes h_\text{mt}, \\
  h_\text{mt} & = c_\text{mt} + \rho_\text{m} c_\text{mm} \otimes
  h_\text{mt} + \rho_\text{t} c_\text{mt} \otimes h_\text{tt}, \\
  h_\text{tt} & = c_\text{tt} + \rho_\text{m} c_\text{mt} \otimes
  h_\text{mt} + \rho_\text{t} c_\text{tt} \otimes h_\text{tt}, \\
  h_\text{mf} & = c_\text{mf} + \rho_\text{m} c_\text{mm} \otimes
  h_\text{mf} + \rho_\text{t} c_\text{mt} \otimes h_\text{tf} +
  \rho_\text{f} c_\text{mf} \otimes h_\text{c}, \\
  h_\text{tf} & = c_\text{tf} + \rho_\text{m} c_\text{mt} \otimes
  h_\text{mf} + \rho_\text{t} c_\text{tt} \otimes h_\text{tf} +
  \rho_\text{f} c_\text{tf} \otimes h_\text{c}, \\
  h_\text{f\/f} & = c_\text{f\/f} + \rho_\text{m} c_\text{mf} \otimes
  h_\text{mf} + \rho_\text{t} c_\text{tf} \otimes h_\text{tf} + \notag
  \\
  & \qquad\qquad\qquad\qquad\quad \rho_\text{f} c_\text{f\/f} \otimes 
  h_\text{f\/f} - \rho_\text{f} c_\text{b} \otimes h_\text{b}, \\
  h_\text{b} & = c_\text{b} + \rho_\text{m} c_\text{mf} \otimes
  h_\text{mf} + \rho_\text{t} c_\text{tf} \otimes h_\text{tf} + \notag
  \\
  & \qquad\qquad\qquad\qquad\quad \rho_\text{f} c_\text{c} \otimes
  h_\text{b} + \rho_\text{f} c_\text{b} \otimes h_\text{c}, \\
  h_\text{c} & = c_\text{c} + \rho_\text{f} c_\text{c} \otimes
  h_\text{c},
\end{align}\end{subequations}
where $\otimes$ represents a convolution in real space. As for the
total correlation functions, one gets $c_\text{f\/f}(r) = c_\text{b}(r)
+ c_\text{c}(r)$. The first three equations are simply the OZ
equations for the bulk binary matrix-template mixture from which the
porous medium is formed, while the others relate to the structure of
the fluid adsorbed in the random solid.

Specialization to the present systems of interest is achieved by
working out the consequences of the identities \eqref{firstidentity}
and \eqref{secondidentity} on the one- and two-body densities. For a
generic PP system, one obtains
\begin{subequations}\begin{gather}
    \rho_\text{t} = \rho_\text{f}, \\
    h_\text{mt}(r) = h_\text{mf}(r), \\
    h_\text{tt}(r) = h_\text{f\/f}(r), \\
    h_\text{tf}(r) = h_\text{b}(r),
\end{gather}\end{subequations}
and equivalently for the direct correlation functions, so that the OZ
equations reduce to
\begin{subequations}\label{OZPP}\begin{align}
    h_\text{mm} & = c_\text{mm} + \rho_\text{m} c_\text{mm} \otimes
    h_\text{mm} + \rho_\text{f} c_\text{mf} \otimes h_\text{mf}, \\
    h_\text{mf} & = c_\text{mf} + \rho_\text{m} c_\text{mm} \otimes
    h_\text{mf} + \rho_\text{f} c_\text{mf} \otimes h_\text{f\/f}, \\
    h_\text{f\/f} & = c_\text{f\/f} + \rho_\text{m} c_\text{mf}
    \otimes
    h_\text{mf} + \rho_\text{f} c_\text{f\/f} \otimes h_\text{f\/f}, \\
    h_\text{b} & = c_\text{b} + \rho_\text{m} c_\text{mf} \otimes
    h_\text{mf} + \rho_\text{f} c_\text{f\/f} \otimes h_\text{f\/f}
    - \rho_\text{f} c_\text{c} \otimes h_\text{c},\\
    h_\text{c} & = c_\text{c} + \rho_\text{f} c_\text{c} \otimes
    h_\text{c}.
\end{align}\end{subequations}

For a RP system with pinning fraction $x$ based on a one-component
bulk fluid with density $\rho$ and total and direct correlation
functions $h(r)$ and $c(r)$, well-known properties of ideal binary
mixtures lead to
\begin{subequations}\begin{gather}
    \rho_\text{m} = x \rho, \\
    \rho_\text{f} = (1-x) \rho, \\
    h_\text{f\/f}(r) = h_\text{mf}(r) = h_\text{mm}(r) = h(r),
\end{gather}\end{subequations}
and analogously for the direct correlation functions, so that the OZ
equations become
\begin{subequations}\label{OZRP}\begin{align}
    h & = c + \rho c \otimes h, \\
    h_\text{b} & = c_\text{b} + \rho c \otimes h - (1-x) \rho
    c_\text{c} \otimes h_\text{c}, \\
    h_\text{c} & = c_\text{c} + (1-x) \rho c_\text{c} \otimes
    h_\text{c}.
\end{align}\end{subequations}

Some features of the above equations are very appealing with a view to
developing integral-equation theoretical approaches. The most obvious
one is their mere number. Indeed, not surprisingly, a significant
reduction in the number of unknown functions and OZ equations linking
them ensues from the special symmetries of the PP and RP systems. It
is particularly drastic in the case of the RP system, where only two
linearly independent OZ equations remain [remember that
$h_\text{f\/f}(r) = h_\text{b}(r) + h_\text{c}(r)$ and
$c_\text{f\/f}(r) = c_\text{b}(r) + c_\text{c}(r)$].  Since one
generically needs at least two of the three functions
$h_\text{f\/f}(r)$, $h_\text{b}(r)$ and $h_\text{c}(r)$ in order to
characterize the pair structure of a fluid in a statistically
homogeneous disordered environment, this is actually the smallest
possible number, whose only known realization so far was in the case
of a Gaussian random field \cite{MenDas94PRL}, i.e., of a
non-particle-based disordered environment. Thanks to this extreme
compactness of the formalism, with its minimal number of unknowns and
coupled equations, one can reasonably expect that the development and
the implementation of advanced integral equation theories for the
present class of systems will be made much easier.

The situation is actually even more favorable than that. Indeed, from
Eqs.~\eqref{OZPP} and \eqref{OZRP}, one can immediately see that the
problem of computing the pair correlations between the different types
of particles present in the system, which amounts to a calculation on
the bulk fluid from which the confined fluid-matrix system is
prepared, is completely separated from the one of obtaining the two
contributions to $h_\text{f\/f}(r)$ which split under the effect of
randomness.  Therefore, since well established integral equation
schemes exist which allow one to successfully tackle the first problem
for a variety of interaction potentials \cite{Cac96PR,macdohansen3ed},
one can build on them, so that the only actual issue for the present
systems of interest boils down to the computation of $h_\text{b}(r)$
and $h_\text{c}(r)$, $h_\text{f\/f}(r)$ being considered as known.

There are different possible ways of taking advantage of this
result. One of them is to note that, because of the separation of the
bulk- and confinement-specific parts of the calculation, one should
not feel bound to treat both problems with the same type of
approximation. Therefore, for a given description of the bulk fluid
system, one can freely experiment with closures for $h_\text{b}(r)$,
$h_\text{c}(r)$, $c_\text{b}(r)$ and $c_\text{c}(r)$.  In combination
with quantitative comparisons with computer simulation data, whose
accumulation is greatly facilitated by the use of
Eqs.~\eqref{firstidentity} and \eqref{secondidentity} (see the
discussion in the previous section), this should lead to insight into
this crucial but not so well understood aspect of the physics of
fluids in random environments, with a real prospect that accurate
theories can be developed, not only for the PP and RP systems, but for
fluids in quenched disorder in general.

\begin{figure}
  \includegraphics*{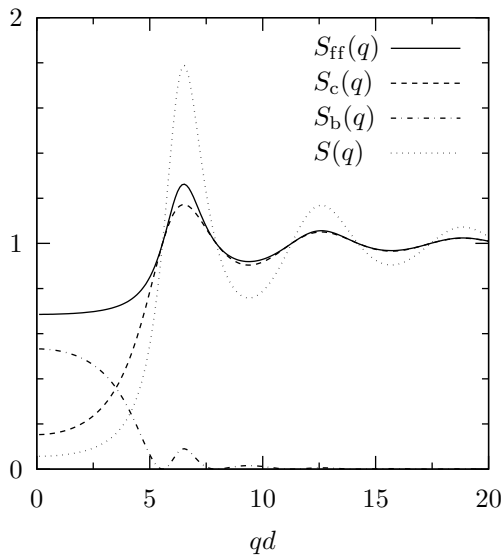}  
  \caption{\label{figPY} Structure factors of a hard-sphere randomly
    pinned fluid-matrix system, obtained analytically in the
    Percus-Yevick approximation.  $S_\text{f\/f}(q)$, $S_\text{c}(q)$,
    and $S_\text{b}(q)$ are, respectively, the fluid-fluid, the
    connected and the blocking structure factors of the confined
    system, $S(q)$ is the structure factor of the bulk fluid from
    which the fluid-matrix system is prepared. The compacity of the
    original bulk system is $\phi = (\pi/6) \rho d^3 = 0.36$, with $d$
    the diameter of the spheres, and the pinning fraction is $x=2/3$.}
\end{figure}

An alternative strategy is to use the same closure approximation for
both parts of the calculation and to concentrate on integral equation
theories that have analytic solutions in the bulk, trying to generate
analytic solutions for the confined system as well. Such solutions are
extremely scarce in the theory of fluids in disordered environments,
even within basic approximation schemes. They might be useful in a
number of different ways. For instance, as demonstrated by many
advances in the case of bulk systems \cite{Cac96PR,macdohansen3ed},
they might represent precious starting points for the development of
more elaborate approximations. Or they might be used to produce at a
minimal computational cost the large amounts of structural data that
are required as input by other types of calculations, such as the
prediction of dynamical phase diagrams in the framework of the
mode-coupling theory \cite{Kra05PRL,Kra05JPCM,Kra07PRE,Kra09PRE}.

It turns out that closures which fall in the class of the so-called
Madden-Glandt approximations, i.e., characterized by an identically
vanishing blocking direct correlation function
\cite{GivSte92JCP,LomGivSteWeiLev93PRE,GivSte94PA}, are particularly
well suited for such an approach. They include the Percus-Yevick (PY)
and mean-spherical (MSA) approximations, for which, precisely, a
number of analytic solutions for bulk systems are well known
\cite{Cac96PR,macdohansen3ed}.  As an illustration of this procedure,
we report in Fig.~\ref{figPY} the structure factors of a hard-sphere
RP system obtained analytically with the PY closure. Since
$c_\text{b}(r) = 0$ and $c_\text{c}(r) = c(r)$ in this approximation,
they are simply expressed in terms of the analytically known Fourier
transform of the PY direct correlation function of the bulk
hard-sphere fluid, $\tilde{c}(q)$ [$\tilde{f}(q)$ denotes the Fourier
transform of $f(r)$], as
\begin{subequations}\begin{align}
    S(q) & = 1+\rho\tilde{h}(q) = \frac{1}{1-\rho\tilde{c}(q)}, \\
    S_\text{f\/f}(q) & = 1+\rho_\text{f}\tilde{h}_\text{f\/f}(q) =
    \frac{1-x\rho\tilde{c}(q)}{1-\rho\tilde{c}(q)}, \\
    S_\text{c}(q) & = 1+\rho_\text{f}\tilde{h}_\text{c}(q) =
    \frac{1}{1-(1-x)\rho\tilde{c}(q)}, \\
    S_\text{b}(q) & = \rho_\text{f}\tilde{h}_\text{b}(q) =
    \frac{x(1-x)\rho^2\tilde{c}(q)^2}
    {[1-\rho\tilde{c}(q)][1-(1-x)\rho\tilde{c}(q)]},
\end{align}\end{subequations}
where $S(q)$ is the structure factor of the bulk fluid from which the
fluid-matrix system is prepared, and $S_\text{f\/f}(q)$,
$S_\text{c}(q)$, and $S_\text{b}(q)$ are, respectively, the
fluid-fluid, the connected and the blocking structure factors of the
confined system.

Finally, beside integral equation theories, other liquid-state
theoretical approaches could either be directly applied to the PP and
RP systems or at least take advantage of their simplifying features in
some parts of their development. For instance, one might turn to
perturbation theory \cite{macdohansen3ed} and, in the same spirit as
in previous work on QA systems \cite{KieRosTarMon97JCP} and fluids in
aerogels \cite{KraKieRosTar01JCP}, study athermal PP or RP reference
systems decorated with attractive fluid-matrix and fluid-fluid
interactions. One would then benefit from all the above results for
the description of the reference system. Note in passing that, while
integral equation theories enforce the complementarity between the
fluid and the matrix at all state points and thus deal with a
temperature-dependent structure of the porous solid, this would not be
the case in such a perturbative scheme, where the disordered matrix
exclusively inherits its properties from the reference athermal
system.

\section{Applications in dynamical studies}

So far, the discussion has essentially been on how the knowledge of
the properties of the bulk fluid on which a PP system is based leads
to considerable simplifications in the study of the confined fluid
system. Then, one might wonder whether, conversely, information on the
latter might help understanding aspects of the physics of the
former. In this section, we show that such opportunities might exist
when dealing with dynamics.

Operationally, any fluid-matrix system in which the porous solid is
represented by fixed randomly placed particles can be described in
dynamical terms as a binary mixture of fluid and matrix particles
with, respectively, finite and infinite masses when the dynamics is
Newtonian, or finite and zero free-diffusion coefficients when it is
Brownian. But, in the case of the PP systems, there is more than
that. Indeed, it is well known that, in classical statistical
mechanics, the configurational properties of a bulk binary mixture are
independent of the finite masses or free-diffusion coefficients of its
two constituents. Equation~\eqref{firstidentity} shows that this
remains true when one mass is sent to infinity or one free-diffusion
coefficient to zero, provided the averaging over the configurations of
the now immobile matrix particles is performed as a disorder average
or, equivalently, as a volume average over a macroscopic sample.  So,
a PP system can effectively be considered as an asymptotic case of a
binary mixture with strong dynamical asymmetry, obtained via a well
defined limiting procedure \cite{VirAraMed95PRL,VirMedAra95PRE,%
  ChaJuaMed08PRE,FenMryPryFol09PRE}. This is a unique situation among
the different particle-based models of disordered porous media, which
originates in the fact that the solid matrix in a PP system is not
prepared independently of the adsorbed fluid, but in its presence in
equilibrium conditions.

We might now try and transport the above limiting procedure into the
realm of dynamical theories. In order to do so, our starting point
will be the description of the relaxation of the collective density
fluctuations in bulk binary mixtures in the framework of the
Mori-Zwanzig formalism \cite{macdohansen3ed}, first for Newtonian
dynamics.

So, we start with a bulk binary mixture with components labeled f and
m. Their particle numbers, number fractions, number densities, and
masses are denoted by $N_\text{f}$ and $N_\text{m}$, $x_\text{f}$ and
$x_\text{m}$, $\rho_\text{f}$ and $\rho_\text{m}$, $m_\text{f}$ and
$m_\text{m}$, respectively. The dynamical variables of interest are
the Fourier components of the microscopic densities at wave vector
$\mathbf{q}$ and time $t$, $\rho_\text{f}(\mathbf{q},t)$ and
$\rho_\text{m}(\mathbf{q},t)$, from which one forms the static
structure factors,
\begin{equation}
  S_{ij}(q) = \frac{\langle \rho_{i}(\mathbf{q},0)
    \rho_{j}(\mathbf{-q},0) \rangle}{N_\text{f}+N_\text{m}} ,
\end{equation}
and density fluctuation autocorrelation functions,
\begin{equation}
  F_{ij}(q,t) = \frac{\langle \rho_{i}(\mathbf{q},t)
    \rho_{j}(\mathbf{-q},0) \rangle}{N_\text{f}+N_\text{m}}.
\end{equation}
Using standard projection-operator methods \cite{macdohansen3ed}, the
latter functions are shown to obey generalized Langevin equations
which read in matrix form
\begin{equation} \label{Newtonian} \ddot{\mathbf{F}}(q,t) +
  \mathbf{\Omega}^2(q) \mathbf{F}(q,t) + \int_0^t d\tau
  \mathbf{M}(q,t-\tau) \dot{\mathbf{F}}(q,\tau)=\mathbf{0},
\end{equation}
with initial conditions $\mathbf{F}(q,0)=\mathbf{S}(q)$ and
$\dot{\mathbf{F}}(q,0)=\mathbf{0}$. $\mathbf{M}(q,t)$ is the matrix of
memory kernels and the frequency matrix $\mathbf{\Omega}^2(q)$ is
given by
\begin{equation} 
  [\mathbf{\Omega}^2]_{ij}(q) = q^2 \frac{k_B T}{m_{i}}
  x_{i} [\mathbf{S}^{-1}]_{ij}(q),
\end{equation}
where $\mathbf{S}^{-1}(q)$ is the matrix inverse of $\mathbf{S}(q)$.

In the limit $m_\text{m}\to+\infty$, which generates a PP system,
$\rho_\text{m}(\mathbf{q},t)$ is not a dynamical variable anymore, so
that $F_\text{mf}(q,t)[ = F_\text{fm}(q,t) ]$ and $F_\text{mm}(q,t)$
remain equal to the corresponding structure factors at all times
\footnote{This results from equalities such as $\overline{\langle
    \rho_\text{f}(\mathbf{q},t) \rho_\text{m}(\mathbf{-q},0) \rangle}
  = \overline{\langle \rho_\text{f}(\mathbf{q},t) \rangle
    \rho_\text{m}(\mathbf{-q},0)} = \overline{\langle
    \rho_\text{f}(\mathbf{q},0) \rangle \rho_\text{m}(\mathbf{-q},0)}
  = \overline{\langle \rho_\text{f}(\mathbf{q},0)
    \rho_\text{m}(\mathbf{-q},0) \rangle} $. }. As a result, the
dynamical equation for $F_\text{f\/f}(q,t)$ obtained by expanding the
matrix products in Eq.~\eqref{Newtonian} reduces to
\begin{multline} 
  \ddot{F}_\text{f\/f}(q,t) + \Omega_\text{f\/f}^2(q)
  F_\text{f\/f}(q,t) + \Omega_\text{fm}^2(q) S_\text{mf}(q) + \\
  \int_0^t d\tau M_\text{f\/f}(q,t-\tau) \dot{F}_\text{f\/f}(q,\tau) =
  0,
\end{multline}
which at infinite time leads to
\begin{equation} \label{infinitetime}
  \Omega_\text{f\/f}^2(q) f_\text{f\/f}(q) + \Omega_\text{fm}^2(q)
  S_\text{mf}(q) + m_\text{f\/f}(q) \left[ f_\text{f\/f}(q) -
    S_\text{f\/f}(q) \right] = 0,
\end{equation}
where
\begin{gather}
  f_\text{f\/f}(q) = \lim_{t\to\infty} \lim_{m_\text{m}\to+\infty}
  F_\text{f\/f}(q,t), \\
  m_\text{f\/f}(q) = \lim_{t\to\infty} \lim_{m_\text{m}\to+\infty}
  M_\text{f\/f}(q,t).
\end{gather}
Note that the order of the limits is essential.

Our goal is to compute $m_\text{f\/f}(q)$, so it remains to evaluate
$f_\text{f\/f}(q)$. If one assumes that the dynamics of the PP system
is ergodic, it is expected on general grounds that
\begin{multline}
  f_\text{f\/f}(q) = \lim_{t\to\infty} \frac{\overline{\langle
      \rho_\text{f}(\mathbf{q},t) \rho_\text{f}(\mathbf{-q},0)
      \rangle}}{N_\text{f}+N_\text{m}} = \\
  \frac{\overline{\langle \rho_\text{f}(\mathbf{q},t) \rangle \langle
      \rho_\text{f}(\mathbf{-q},0) \rangle}}{N_\text{f}+N_\text{m}} =
  x_\text{f} \rho_\text{f} \tilde{h}_\text{b}(q),
\end{multline}
while $S_\text{f\/f}(q) = x_\text{f} [1 + \rho_\text{f}
\tilde{h}_\text{f\/f}(q)]$. Note that, compared to the previous
section, additional factors $x_\text{f}$ appear due to the use of a
different normalization for the structure factors. By combining
Eq.~\eqref{infinitetime} with the set of OZ equations \eqref{OZPP}, it
then results that
\begin{equation} \label{memory} m_\text{f\/f}(q) = q^2 \frac{k_B
    T}{m_\text{f}} \rho_\text{f} \tilde{c}_\text{b}(q).
\end{equation}

For Brownian dynamics with free-diffusion coefficients $D^0_\text{f}$
and $D^0_\text{m}$, one gets, instead of Eq.~\eqref{Newtonian}, the
matrix equation \cite{NagBerDho99JCP}
\begin{equation}
  \dot{\mathbf{F}}(q,t) + \mathbf{\Omega}(q)
  \mathbf{F}(q,t) + \int_0^t d\tau \mathbf{M}(q,t-\tau)
  \dot{\mathbf{F}}(q,\tau)=\mathbf{0},
\end{equation}
where $\mathbf{M}(q,t)$ is now the matrix of irreducible memory
functions and the damping matrix $\mathbf{\Omega}(q)$ is given by
\begin{equation} 
  \Omega_{ij}(q) = q^2 D^0_{i} x_{i} [\mathbf{S}^{-1}]_{ij}(q).
\end{equation}
Repeating the above steps with $D^0_\text{m}\to0$, one obtains
\begin{equation} \label{memory2} m_\text{f\/f}(q) = q^2 D^0_\text{f}
  \rho_\text{f} \tilde{c}_\text{b}(q).
\end{equation}

Equations \eqref{memory} and \eqref{memory2} represent bridges between
the statics of the PP systems and the dynamics of the fully annealed
mixtures with a strong dynamical asymmetry. We suggest that they could
be usefully incorporated into the development of phenomenological
expressions for the memory kernel $M_\text{f\/f}(q,t)$ reflecting the
separation of time scales between the short-time relaxation due to the
motion of the fast particles constrained by the nearly immobile slow
ones and the long-time relaxation associated with the ultimate
decorrelation of the positions of the latter. For instance, since a
number of observables in dynamically asymmetric mixtures show features
reminiscent of the cage effect and two-step relaxation
\cite{FenMryPryFol09PRE}, one could imagine building two-step memory
kernels with a plateau value around the above determined
$m_\text{f\/f}(q)$.

How this should be done in practice will depend on the details of the
dynamical theory at hand \cite{macdohansen3ed,boonyip} and is beyond
the scope of the present paper. At this point, however, a few words of
caution might be in order. First, the above argument specifically
deals with one possible source of dynamical asymmetry, namely a wide
separation in the kinetic parameters (masses or free-diffusion
coefficients) of the mixture components
\cite{VirAraMed95PRL,VirMedAra95PRE,ChaJuaMed08PRE,FenMryPryFol09PRE}.
It is not clear if and how it can be transferred to cases where the
dynamical asymmetry has another origin, such as size disparity in
dense systems (for recent studies, see
Refs.~\cite{MorCol06PRE,VoiHor09PRL}). Second, it is already known
that not all theoretical frameworks are able to bridge between
annealed mixtures and fluid-matrix systems. A prominent example is the
mode-coupling theory \cite{Kra05PRL,Kra05JPCM,Kra07PRE,Kra09PRE} whose
equations for these two problems have been shown to be generically
incompatible \footnote{Following Ref.~\cite{Kra07PRE}, the likely
  reason for this is that the derivation of the mode-coupling theory
  for fluid-matrix systems critically relies on the distinction
  between connected and blocking correlation functions that does not
  make sense for bulk systems.}. In such a case, the connection
described above does not bring any additional insight, neither into
the dynamics of asymmetric bulk mixtures nor into that of RP and PP
systems, and each problem has to be analyzed separately within the
chosen theoretical framework.

Obviously, and independently of the connection suggested above, the
dynamics of the RP and PP systems has an interest of its own, as a
simple model of a fluid in a disordered confining environment. In
particular, because the computation of static averages is made easy
and well controlled in these systems thanks to the identities of
Sec.~\ref{identities}, they are very appealing models to test
dynamical theories which, quite generally and as illustrated by the
mode-coupling theory \cite{Kra05PRL,Kra05JPCM,Kra07PRE,Kra09PRE},
require prior knowledge of a number of static properties to deliver
their predictions. We stress here that the mode-coupling theory for
fluids in random porous matrices
\cite{Kra05PRL,Kra05JPCM,Kra07PRE,Kra09PRE} does not make any
assumption on the statistics of the disordered solid (beyond
statistical homogeneity) and thus applies as it stands to RP and PP
systems.

These models can also be put to good use to investigate the
configurational consequences of permanent trapping in systems with
hard-core fluid-matrix interactions. Indeed, it is well known that,
for such systems at any nonvanishing matrix density, there is always a
finite probability that particles are trapped in finite domains
disconnected from the rest of the matrix void space
\cite{KerMet83JPA,KamHofFra08EL}. This trapping phenomenon culminates
in a diffusion-localization transition which coincides with the
percolation transition marking the point after which the matrix void
space only consists of finite disconnected domains
\cite{HofFraFre06PRL,HofMunFreFra08JCP,KamHofFra08EL,%
  SunYet06PRL,SunYet08JPCB,SunYet08JCP,BabGimNic08JPCB}.  Thus, the
exploration of the volume accessible to the fluid is never completely
ergodic and one expects to see at least quantitative differences in
configurational properties depending on whether the thermal averaging
is performed as an ensemble or a trajectory average, denoted by
$\langle \cdots \rangle$ and $[\cdots]$, respectively.  For a generic
fluid-matrix system, such an investigation is hampered by a number of
technical difficulties. For instance, if double averages of the form
$\overline{\langle A \rangle}$ and $\overline{[ A ]}$ are to be
compared, the estimation of the former requires the use of smart
simulation methods (typically, with nonlocal updates) in order to
overcome the lack of ergodicity of the system. In the case of
quantities such as $\overline{\langle A \rangle \langle B \rangle}$
and $\overline{[ A ][ B ]}$, one has in addition to face the issues
mentioned at the end of Sec.~\ref{identities} for the computation of
such correlation functions. Tackling the problem with a RP or a PP
system does not eliminate all difficulties, but significant
simplifications occur which should make the various computations much
more accessible. Thus, thanks to Eq.~\eqref{firstidentity}, the double
average $\overline{\langle A \rangle}$ can be efficiently retrieved
from studies of the bulk fluid system on which the fluid-matrix model
is based, in which no trapping occurs, hence with no need for special
sampling techniques.  Also, rather than computing $\overline{\langle A
  \rangle \langle B \rangle}$ and $\overline{[ A ][ B ]}$, one can
follow Eq.~\eqref{secondidentity} and concentrate on correlation
functions involving the positions of the fluid particles at the time
of the preparation of the matrix, which are much easier to compute and
have essentially the same physical content.

\section{Conclusion}

In this paper, the statistical properties of the partly pinned fluid
systems, a special class of models of fluids confined in disordered
porous matrices, have been studied. These systems, which in most
respects are just typical examples of fluid-matrix models, are singled
out by a peculiar complementarity between the mobile and immobile
components, which originates from the fact that the confining random
solid is prepared in presence of and in equilibrium with the adsorbed
fluid. A special symmetry results, with major repercussions on the
configurational properties of the system. In particular, simple
identities hold, which relate different types of configurational
averages, either relative to the fluid-matrix system or to the bulk
fluid from which it is prepared.

By taking advantage of these identities in computer simulation
studies, it seems that, with these systems, interesting opportunities
are opening up for efficient and accurate computations of quantities
that are usually quite difficult to estimate in generic fluid-matrix
models, in particular the crucially important connected and
disconnected correlation functions. In our opinion, this should make
the PP fluid systems the models of choice for a number of theoretical
developments in the field of fluids adsorbed in disordered porous
solids. Indeed, in many circumstances, either theories aim at
predicting structural properties, as do integral equation theories,
and then it is clearly an advantage that their predictions can be
readily tested against high-quality simulation data, or they take
structural data as input in order to predict other physical
properties, such as the dynamics, and then it is extremely convenient
that this input can be easily and accurately determined. In all cases,
one can reasonably expect that the insight gained on the PP systems
will be transferable to other types of fluid-matrix models, so that
global progress can be made in our understanding of fluids in random
environments.

There is however a price to pay for these conveniences. Indeed, by
construction of the model, one cannot change the parameters describing
the adsorbed fluid without a similar and complementary modification of
the matrix. This is clearly at odd with the typical experimental
situation, where, usually, the confining conditions are given and
fixed and the fluid characteristics can be altered freely and
independently, and might lead to difficulties of interpretation, when
evolutions of the properties of the system across its parameter space
are considered.

Finally, some extensions of the present work could be of interest. One
obvious direction is towards heterogeneous partly pinned fluid
systems, in which all the particles of a bulk system located in a
predefined spatial domain are stopped, leaving a fluid confined in a
pore or in contact with a solid interface. This setup has been
introduced in order to study various aspects of glass formation
\cite{SchKobBin04JPCB,BouBir04JCP,CavGriVer07PRL,BirBouCavGriVer08NatPhys}
and an identity analogous to Eq.~\eqref{firstidentity} is known to
hold \cite{SchKobBin04JPCB}. Note however that, in order to prevent
the invasion of the solid by the fluid, the authors of
Ref.~\cite{SchKobBin04JPCB} mention the need to introduce a hard
separation at the fluid-matrix interface which might lead to
complications and approximations absent in the homogeneous
case. Another possibility would be to translate the problem in the
language of magnetism, where the pinning process would operate on spin
variables. This would generate examples of spin systems with
correlated random field and/or dilution for which suitable adaptations
of Eqs.~\eqref{firstidentity} and \eqref{secondidentity} would apply.

%\bibliography{../../Bibtex/abbrev,../../Bibtex/confinedfluids_theo,%
%../../Bibtex/confinedfluids_exp}


\end{document}